\documentclass[aps,twocolumn,groupeaddress,prb]{revtex4}

\newcommand{\me}{\mathrm{e}}

\newcommand{\dif}{\mathrm{d}}

\usepackage{graphicx}

\usepackage{amsmath}

\usepackage{amsfonts}

\usepackage{dcolumn}% Align table columns on decimal point

\usepackage{bm}% bold math

\begin{document}

\preprint{LA-UR-06-5877}

\title{Balancing Local Order
and Long-Ranged Interactions in the Molecular Theory of Liquid Water}% Force line breaks with \\

\author{J. K. Shah}\altaffiliation{Department of Chemical and Biomolecular
Engineering, Ohio State University, Columbus OH 43210 USA}

\author{D. Asthagiri}
\altaffiliation{Department of Chemical and Biomolecular Engineering,
Johns Hopkins University, Baltimore MD 21218 USA}

\author{L. R. Pratt}
\altaffiliation{Theoretical Division, Los Alamos National Laboratory,
Los Alamos NM 87545 USA}

\author{M. E. Paulaitis$^\ast$}

\date{\today}
             
\begin{abstract} 
A molecular theory of liquid water is identified and studied on the
basis of computer simulation of the TIP3P model of liquid water. This
theory would be exact for models of liquid water in which
the intermolecular interactions vanish outside a finite spatial range,
and therefore provides a precise analysis tool for investigating the
effects of longer-ranged intermolecular interactions.  We show how local
order can be introduced through quasi-chemical theory.  Long-ranged
interactions are characterized generally by a conditional distribution
of binding energies, and this formulation  is interpreted as a
regularization of the primitive statistical thermodynamic problem. These
binding-energy distributions for liquid water are observed to be
unimodal. The gaussian approximation proposed is remarkably successful
in predicting the Gibbs free energy and the molar entropy of liquid
water, as judged by comparison with numerically exact results. The
remaining discrepancies are subtle quantitative problems that do have
significant consequences for the thermodynamic properties that
distinguish water from many other liquids.  The basic subtlety of liquid
water is found then in the \emph{competition} of several effects which
must be quantitatively balanced for realistic results.
\end{abstract}

\maketitle

\section*{Introduction}

\begin{figure}
    \includegraphics[width=3.2in]{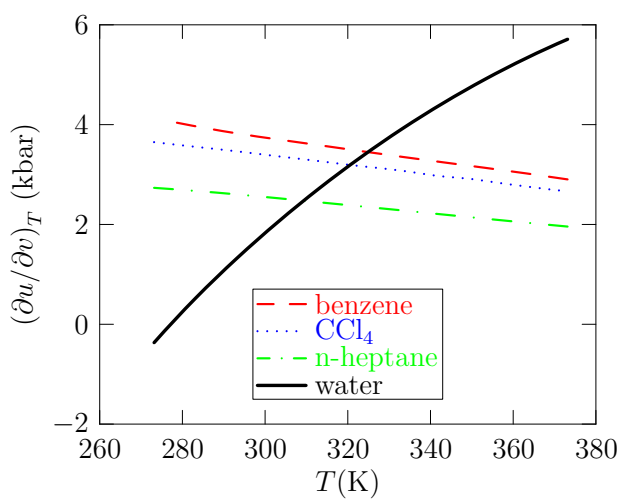}
    \caption{The  internal pressure of several solvents
    as a function of temperature along their vapor saturation curve.\cite{Rowlinson}  For
    van der Waals liquids $\left(\partial u/\partial v\right)_T \approx
    a \rho^2$.  Organic solvents conform to this expectation, but water 
    is qualitatively different.\label{fig:internal_pressure}}  
\end{figure}

On the basis of the temperature dependence of its macroscopic
properties, liquid water belongs to a class of liquids often referred to
as ``associated liquids.'' Fig.~\ref{fig:internal_pressure} gives one
example of a characteristic distinction between liquid water and common
organic solvents: the temperature dependence of the internal pressure.
For van der Waals liquids, this temperature dependence is related
directly to the temperature dependence of the liquid density. Organic
solvents conform to this expectation, but liquid water exhibits a
contrasting behavior, and that alternative behavior is attributed
typically to the temperature dependence of water association through
hydrogen bonding interactions.

As an associated liquid, water is  regarded broadly also as a network
liquid, though a definition of ``network liquid'' is usually not
attempted. One aspect of the network view of liquid water  is that
models with defined short-ranged hydrogen bonding interactions that
vanish precisely outside a finite range are designed specifically to
study the network characteristics of that fluid.  The work of Peery \&
Evans\cite{PeeryTB:Assfwf} gives a recent example that reviews and
advances the forefront of statistical mechanical theory built explicitly
on that network concept; this line of inquiry does indeed have an
extended history, and the citations
\cite{ANDERSENHC:CLUEFH1,ANDERSENHC:CLUEFH2,DAHLLW:Atat,AranovichG:Almf,%
ECONOMOUIG:Cheqap} give further examples.  The finite range of interactions is a common
feature of lattice-gas models of liquid
water.\cite{FLEMINGPD:I,FLEMINGPD:II,StillingerFH:Theamm}

The theory investigated here is motivated by a simple, surprising
result which can be regarded as a theorem for models with intermolecular
interactions that vanish outside a radius $\lambda$, whether or not the liquid  under consideration
would be regarded as a network liquid on some other basis. 
Specifically, it is that the excess chemical potential is precisely
\begin{eqnarray}
\beta \mu^\mathrm{{ex}} = \beta \mu^\mathrm{{ex}}_\mathrm{{HS}} \left(\lambda\right)
+  \ln p\left(n_\lambda = 0\right) ~.
\label{eq:theorem}
\end{eqnarray}
This excess chemical potential $\mu^\mathrm{{ex}}$ is the Gibbs free
energy per mole, in excess of the ideal contribution  at the same density
and temperature, of the one-component fluid considered.  The
thermodynamic temperature is $T=1/k\beta$ where $k$ is the Boltzmann's
constant, $\mu^\mathrm{{ex}}_\mathrm{{HS}}$ is the excess chemical
potential for a hard-sphere solute, with $\lambda$ the distance of
closest approach, at infinite dilution in the fluid.  In the final term,
$p\left(n_\lambda = 0\right) $ is the probability that a specific actual
molecule in the fluid has zero (0) neighbors within the radius
$\lambda$.

After articulation, the theorem Eq.~\ref{eq:theorem} is
straightforwardly understood, and a simple proof is given below. As an
orienting comment, we note that Eq.~\ref{eq:theorem} has been useful
already in the theory of the hard-sphere fluid.\cite{Beck:2006} Of
course, hard-sphere interactions are not involved literally in the
interactions of physical systems, and it will transpire below that the
hard-sphere contribution in  Eq.~\ref{eq:theorem} simply serves to
organize the statistical information, not as an assumed feature of an
underlying intermolecular potential model.

We emphasize that Eq.~\ref{eq:theorem} is correct if intermolecular
interactions vanish beyond the radius $\lambda$, independently of whether
the liquid under consideration is a network liquid.  Real network
liquids have strong, short-ranged binding interactions, but invariably
longer-ranged interactions as well.  This   paper investigates the
simplest approximate extension of the Eq.~\ref{eq:theorem} to real
liquids, an extension which is realizable, and which is conceived on the basis of
observable data, not from consideration of an assumed interaction model.
 The specific issues we address quantitatively include whether the
effects of longer-than-near-neighbor interactions are simple for
realistic models of liquid water, and whether reasonable values of
$\lambda$ can be found that make the statistical theory of liquid water
particularly simple with realistic intermolecular interactions.

\section*{Statistical Thermodynamic Theory}

We seek $\mu^\mathrm{{ex}}$ on the basis of accessible simulation data.
Since Eq.~\ref{eq:theorem} contains $\mu^\mathrm{{ex}}_\mathrm{{HS}}$,
we consider evaluating that quantity on the basis of simulation of the
realistically modeled liquid of interest. For that purpose, the
potential distribution theorem (PDT)
\cite{CF4,Beck:2006} difference formula yields
\begin{eqnarray}
e^{-\beta(\mu^\mathrm{{ex}}_\mathrm{{HS}} - \mu^\mathrm{{ex}})} = 
p\left(n_\lambda = 0\right) 
\int\limits_{-\infty}^{+\infty} P\left(\varepsilon \vert n_\lambda = 0\right) 
e^{\beta\varepsilon} \dif\varepsilon ~.
\label{eq:derivation}
\end{eqnarray}
If we take  $\mu^\mathrm{{ex}}_\mathrm{{HS}}$ as known, \cite{Ashbaugh:2006} Eq.~\eqref{eq:derivation} is an equation for
$\mu^\mathrm{{ex}}$.  $P\left(\varepsilon \vert n_\lambda = 0\right) $
is the conditional probability of the binding energy of a specific
molecule for the event that $ n_\lambda = 0$, \emph{i.e.,} the specific
molecule considered has zero (0) neighbors within radius $\lambda$. 
$p\left(n_\lambda = 0\right)$ is the marginal probability of that event. 
If the intermolecular interactions vanish for ranges as large as
$\lambda$, those binding energies are always zero (0).  This yields
Eq.~\ref{eq:theorem} under the assumptions noted.

For general  interactions, we regard the conditioning of
Eq.~\ref{eq:derivation} when $\lambda >0$ as a {\it regularization} of
the statistical problem embodied in Eq.~\eqref{eq:derivation} when
$\lambda=0$, which is impossible on the basis of a direct calculation.
After regularization, the statistical problem becomes merely difficult.
For $\lambda\rightarrow\infty$, a gaussian distribution model for
$P\left(\varepsilon\vert n_\lambda = 0 \right)$ should be accurate since
then many solution elements will make small, weakly-correlated
contributions to $\varepsilon$.  The marginal probability
$p\left(n_\lambda = 0\right)$ becomes increasingly difficult to evaluate
as $\lambda$ becomes large, however. For $\lambda$ on the order of
molecular length scales typical of dense liquids, a simple gaussian
model would accept some approximation error as the price for manageable
statistical error. If $P\left(\varepsilon\vert n_\lambda = 0\right)$ is
modeled by a gaussian of mean $\left<\varepsilon\vert n_\lambda =
0\right>$ and variance $\left < \delta \varepsilon ^2 \vert n_\lambda =
0\right >$, then
\begin{multline}
\mu^\mathrm{{ex}} - \mu^\mathrm{{ex}}_\mathrm{{HS}} \left(\lambda\right)
- kT \ln p\left(n_\lambda = 0\right) 
-  \left<\varepsilon\vert n_\lambda = 0\right> \\ 
= 
\frac{1}{2 kT}\left < \delta \varepsilon ^2 \vert n_\lambda = 0\right >
~.
\label{eq:working_eq_gaussian}
\end{multline}
We regard the formulation Eq.~\ref{eq:working_eq_gaussian} merely as a
parsimonious use of statistical information that might be obtained from
a simulation record.  This simple  model motivates the following analyses.

An essentially thermodynamic derivation of Eq.~\ref{eq:derivation}, one that
illustrates the connection with quasi-chemical theory, was given
previously.\cite{pratt07}  Our result there was formulated as
\begin{eqnarray}
\beta \mu_\alpha^\mathrm{ex} = \beta \mu_{\alpha\mathrm{w}_{m=0}}^\mathrm{ex} - \ln
\left( 1
+ \sum_{m\ge 1} K_m\rho_\mathrm{w}{}^m\right)
\label{eq:thermoqct}
\end{eqnarray}
where $\alpha$ indicates the distinguished molecule and `w' the
molecules of the solvent medium, for example water.  Comparison of
Eqs.~\ref{eq:derivation} and \ref{eq:thermoqct} then suggests the
identification 
\begin{eqnarray}
p\left( n_\lambda = 0\right) =  \frac{1}
{ 1
+ \sum_{m\ge 1} K_m\rho_\mathrm{w}{}^m}~.
\label{eq:c-contribution}
\end{eqnarray}
Thus the marginal probability of the event ($n_\lambda$ = 0) directly
interrogates chemical contributions involving the chemical equilibrium
ratios $K_n$'s on the basis of the assumed forcefield.  This
correspondence is indeed a basic result of constructive derivations  of
quasi-chemical theory.\cite{Paulaitis:2002} Given adequate simulation
data, explicit evaluation of individual $K_n$'s is not required. 
But more basic chemical analysis of the
$K_n$'s, \emph{e.g.,}
using electronic structure method,\cite{MartinRL:Hydfiw,Rempe99,RempeSB:ThehnN,RempeSB:ThehnL,%
AsthagiriD:Hydsaf,AsthagiriD:Abshfe,AsthagiriD:QuasB2,AsthagiriD:ThehsH,RempeSB:Innsda} is
not implemented either.  We add
for completeness that previous discussions denoted $p\left( n_\lambda =
0\right) = x_0$, with $x_n \propto K_n\rho_\mathrm{w}{}^n$ the
probability that the distinguished molecule has a coordination number of
$n$, and $K_0\equiv$1.\cite{PrattLR:Quatal,Rempe99,Paulaitis:2002,Beck:2006,pratt07}

These equations correctly suggest that this formulation is fully general
for the classical statistical thermodynamic problem considered. This
formulation doesn't make a lattice structural idealization, doesn't
assume  that the interaction potential energy is pairwise decomposable,
doesn't assume that the interaction potential energy model provides a
handy definition of `H-bonded,' and the coordination definition ---
the occupancy within a radius $\lambda$ --- is observational on the basis
of molecular structure.  Effective values of $\lambda$ should  be
established eventually on the basis of the statistical information
secured.

Of course, the chemical potential provides the partial molar entropy
\begin{eqnarray}
\left(\frac{\partial S}{\partial n_\alpha}\right)_{T,p}
= \frac{1}{T}\left(\frac{\partial \left\langle E\right\rangle}{\partial n_\alpha}\right)_{T,p}
+ \frac{p}{T}\left(\frac{\partial \left\langle V\right\rangle}{\partial n_\alpha}\right)_{T,p}
- \frac{\mu_\alpha}{T},
\end{eqnarray}
or for the one-component system considered here
\begin{eqnarray}
\frac{ S^{\mathrm{ex}}}{ nk}
= \frac{\left\langle \varepsilon\right\rangle}{2	k T}
+ \left(\frac{p}{ \rho kT }-1\right)
- \frac{\mu^{\mathrm{ex}}}{kT}~.
\label{S:eq}
\end{eqnarray}
On the right side the quantities  $\left\langle \varepsilon\right\rangle$, and 
$p$ (the fluid pressure) are mechanical properties which are 
directly available from a simulation record.

It is an important but  a tangential point that  the distributions
$P\left(\varepsilon\vert n_\lambda = 0 \right)$ to which this theory
directs our attention are different from the $p(v)$ that have been
frequently considered following the work of Rahman \&
Stillinger;\cite{RAHMANA:Moldsl} see also the review of
Stillinger.\cite{STILLINGERFH:Watr}
$p(v)$ is the distribution with the interpretation that
$\int_{v_1}^{v_2} p(v)\dif v = n (v_2,v_1)$ is the \emph{number} of
molecules neighboring a particular one with pair interaction $v$ in the
range $(v_2,v_1)$. $p(v)$ has been used to convey reasonable definitions
of `H-bonded' on the basis of simulation models of water, and it is
well-designed for the purpose. It doesn't supply the correlations that
the distributions $P\left(\varepsilon\vert n_\lambda = 0 \right)$ use to
determine the free energy.

\section*{Simulation Data}

To test these ideas, 50000 snapshots from simulations\cite{PaliwalA.:Anamp,Paliwal:thesis} performed at 298,
350 and 400 K  were analyzed.
Calculations of $\mu^\mathrm{{ex}}_\mathrm{{HS}}$ and $p(n_\lambda = 0)$
were carried out as described by Paliwal, \emph{et
al.}\cite{PaliwalA.:Anamp} The interaction
energy of a water molecule was calculated as the sum of pairwise
additive van der Waals and electrostatic energies. Lennard-Jones
interactions were considered within 13~\AA, and standard long range
corrections were applied beyond this distance. Electrostatic
interactions were evaluated using the Ewald summation method.

Statistical uncertainties in $\mu^{\mathrm{ex}}$  were computed by
dividing the total number of snapshots into five blocks and evaluating
block averages. For comparison, $\mu^{\mathrm{ex}}$  was determined from
the histogram overlap method after evaluating also the uncoupled
binding-energy  distribution $P^{(0)}\left(\varepsilon\right)$,\cite{Beck:2006} which was obtained by placing 5000 randomly oriented
water molecules at randomly chosen locations in each snapshot.
Table~\ref{tab:results} collects the individual terms for the gaussian
model, Eq.~\eqref{eq:working_eq_gaussian},  at each temperature, and
additionally the entropy evaluated from  Eq.~\ref{S:eq}. The observed
dependence on $\lambda$ of the free energy at each temperature is shown
in   Fig~\ref{fig:murdep}.

\begin{table*}
\begin{center}
\caption{Free energy contributions in kcal/mol associated with the
gaussian model.  The bottom value of each temperature set gives the
corresponding free energy evaluated by the histogram overlap method. 
The rightmost column is the excess entropy per particle.  Note that the
entropy here raises the net free energy (see Eq.~\ref{S:eq}), and the
magnitude of the entropy contribution $TS^{\mathrm{ex}}/n$ ranges here
from 60-75\% of the magnitude of the free energy
$\mu^{\mathrm{ex}}$.\label{tab:results}}

\begin{tabular}{|c|l|ccrcl|r|}
\hline
	   T(K)  
	&  $\lambda$ (nm)  
	&  $\mu^{\mathrm{ex}}_{\mathrm{HS}}\left(\lambda \right)$ 
	&  $+ kT\ln p\left(n_\lambda = 0 \right)$
	&  $+ \left\langle \varepsilon\vert n_\lambda = 0\right\rangle$ 
	&  $+\left\langle \delta \varepsilon^2\vert n_\lambda = 0\right\rangle/2kT$  
	&  $= \mu^{\mathrm{ex}}$   
	& $S^{\mathrm{ex}}/n k$\\ \hline
298 & 0.2600 & 2.93 & -0.04 & -19.74  & 9.87 & -6.98 $\pm$ 0.02 & -5.88
\\
    & 0.2650 & 3.15 & -0.13 & -19.68  & 9.93 & -6.73 $\pm$ 0.03 & -6.30
\\
    & 0.2675 & 3.25 & -0.20 & -19.59  & 9.97 & -6.57 $\pm$ 0.04  & -6.57
\\
    & 0.2700 & 3.35 & -0.31 & -19.46  & 9.98 & -6.44 $\pm$ 0.03 &  -6.79
\\
    & 0.2725 & 3.45 & -0.44 & -19.27  & 9.97 & -6.29 $\pm$ 0.03 & -7.05
\\
    & 0.2750 & 3.56 & -0.60 & -19.03  & 9.92 & -6.15 $\pm$ 0.03 &  -7.28
\\
    & 0.2775 & 3.67 & -0.78 & -18.73  & 9.83 & -6.01 $\pm$ 0.04 & -7.52
\\
    & 0.2800 & 3.78 & -0.98 & -18.39  & 9.71 & -5.88 $\pm$ 0.03 & -7.74
\\
    & 0.2900 & 4.23 & -1.96 & -16.75  & 8.89 & -5.59 $\pm$ 0.04 & -8.23
\\
    & 0.3000 & 4.71 & -3.09 & -14.93  & 7.77 & -5.54 $\pm$ 0.06 &  -8.31
\\
    & 0.3100 & 5.22 & -4.27 & -13.21  & 6.67 & -5.59 $\pm$ 0.18 & -8.23
\\
    & 0.3200 & 5.76 & -5.45 & -11.65  & 5.54 & -5.80 $\pm$ 0.37 & -7.87
\\
    & 0.3300 & 6.32 & -6.64 & -10.30  & 4.78 & -5.84 $\pm$ 0.97 & -7.81\\ 
    &        &      &       &         &      &     -6.49     & -6.71   \\ \hline

350 & 0.2600 & 3.06 & -0.05 & -18.43 & 9.44 & -5.98 $\pm$ 0.02 & -5.65
\\
    & 0.2625 & 3.15 & -0.09 & -18.41 & 9.46 & -5.89 $\pm$ 0.02  & -5.78
\\
    & 0.2700 & 3.45 & -0.33 & -18.14 & 9.47 & -5.55 $\pm$ 0.02 &  -6.27
\\
    & 0.2750 & 3.66 & -0.62 & -17.73 & 9.35 & -5.34 $\pm$ 0.02 & -6.57
\\
    & 0.2800 & 3.87 & -0.99 & -17.15 & 9.10 & -5.17 $\pm$ 0.01  & -6.82
\\
    & 0.2900 & 4.32 & -1.92 & -15.67 & 8.24 & -5.03 $\pm$ 0.04 & -7.02
\\
    & 0.3000 & 4.80 & -3.00 & -14.05 & 7.20 & -5.05 $\pm$ 0.06 &  -6.99
\\
    & 0.3100 & 5.29 & -4.13 & -12.48 & 6.14 & -5.18 $\pm$ 0.10  & -6.80
\\
    & 0.3200 & 5.81 & -5.28 & -11.01 & 5.24 & -5.24 $\pm$ 0.22 &  -6.72
\\
    & 0.3300 & 6.36 & -6.41 &  -9.74 & 4.47 & -5.32 $\pm$ 0.45 &-6.60
\\ 
    &        &      &       &        &      & -5.83       & -5.87      \\ \hline

400 & 0.2600 & 3.03 & -0.06 & -17.19 & 8.97  & -5.25 $\pm$ 0.02 & -5.21
\\
    & 0.2625 & 3.13 & -0.10 & -17.16 & 8.99  & -5.14 $\pm$ 0.03 & -5.35
\\
    & 0.2700 & 3.41 & -0.35 & -16.89 & 8.95  & -4.88 $\pm$ 0.02 & -5.67
\\
    & 0.2750 & 3.61 & -0.63 & -16.49 & 8.80  & -4.71 $\pm$ 0.03 & -5.89
\\
    & 0.2800 & 3.81 & -0.98 & -15.96 & 8.52  & -4.61 $\pm$ 0.03 & -6.01
\\
    & 0.2900 & 4.23 & -1.87 & -14.60 & 7.69  & -4.55 $\pm$ 0.05 & -6.09
\\
    & 0.3000 & 4.68 & -2.89 & -13.10 & 6.72  & -4.59 $\pm$ 0.05 & -6.04
\\
    & 0.3100 & 5.14 & -3.97 & -11.64 & 5.78  & -4.69 $\pm$ 0.06 & -5.91
\\
    & 0.3200 & 5.62 & -5.06 & -10.31 & 4.95  & -4.80 $\pm$ 0.14 & -5.77
\\
    & 0.3300 & 6.13 & -6.12 & -9.12  & 4.27  & -4.84 $\pm$ 0.18 & -5.72\\ 
    &        &      &       &        &       &     -5.31   & -5.13   \\ \hline

\end{tabular}
\end{center}
\end{table*}

\section*{Results} 
\begin{figure}[t]
\includegraphics[width=3.2in]{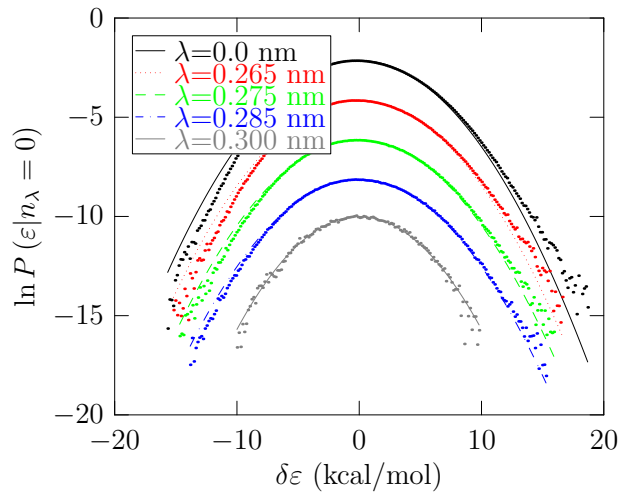}
\caption{Probability density $P(\varepsilon \vert n_\lambda = 0)$ of the
binding energy of a water molecule in liquid water at 298 K. 
$\delta\varepsilon \equiv$ $\varepsilon - \left\langle \varepsilon \vert
n_\lambda = 0\right\rangle$, $\lambda$ =  0.0, \ldots 0.30~nm, from  top
to bottom with successive results shifted incrementally downward by 2
for clarity. The solid lines are the gaussian model for each data
set.\label{Fig:comparison_rcut_full_T300} }
\end{figure}

Fig.~\ref{Fig:comparison_rcut_full_T300} shows that the unconditioned
distribution $P\left(\varepsilon\right)$ displays positive skew, but the
conditioning diminishes that skew perceptibly, as expected.
$P\left(\varepsilon\vert n_\lambda = 0\right)$ is least skewed for the
largest $\lambda$.  All of these results are qualitatively unlike the
binding-energy distributions obtained for a two dimensional model of
liquid water \cite{AndaloroG.:MonCsh} which has been studied further.\cite{HaymetADJ:Hydr'e,SilversteinKAT:Asmw} That previous result is
multi-modal, behavior not seen here.

The distributions  $P\left(\varepsilon\vert n_\lambda =
0 \right)$ shown in Fig.~\ref{Fig:comparison_rcut_full_T300} are also
qualitatively unlike the pair
interaction distributions $p(v)$ \cite{RAHMANA:Moldsl,STILLINGERFH:Watr}
which are different, as was discussed above.

\begin{figure}[b]
\hspace{0.08in}\includegraphics[width=3.1in]{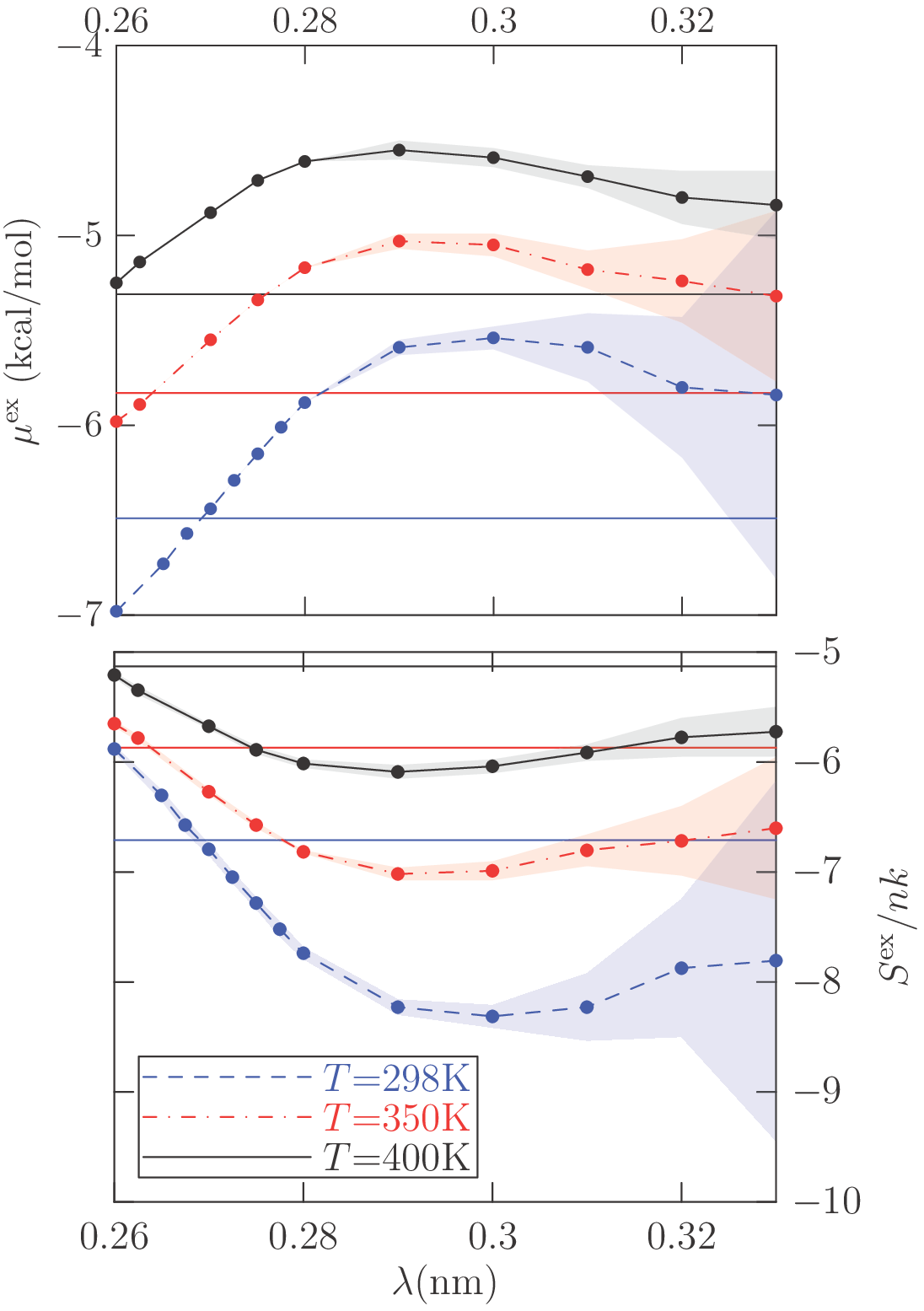}
\caption{upper panel:  Dependence of the free energy $\mu^{\mathrm{ex}}$
predicted by the gaussian model on the conditioning radius $\lambda$.
The horizontal lines are the numerically exact results. The shaded 
areas indicate approximate 95\% confidenc\label{fig:murdep}}
\end{figure}

The conditioning affects both the high-$\varepsilon$ and
low-$\varepsilon$ tails of these distributions.  The mean binding energy
$\left\langle \varepsilon\vert n_\lambda = 0 \right\rangle$ increases with
increasing $\lambda$ [Table~\ref{tab:results}], so we conclude that the
conditioning eliminates  low-$\varepsilon$
configurations more than high-$\varepsilon$ configurations that reflect
less favorable interactions.   Nevertheless, because of the exponential
weighting of the integrand of Eq.~\eqref{eq:derivation} and the 
large variances, the high-$\varepsilon$ side of the distributions
is overwhelmingly the more significant in this free energy prediction.

\begin{figure}[t]
\includegraphics[width=3.2in]{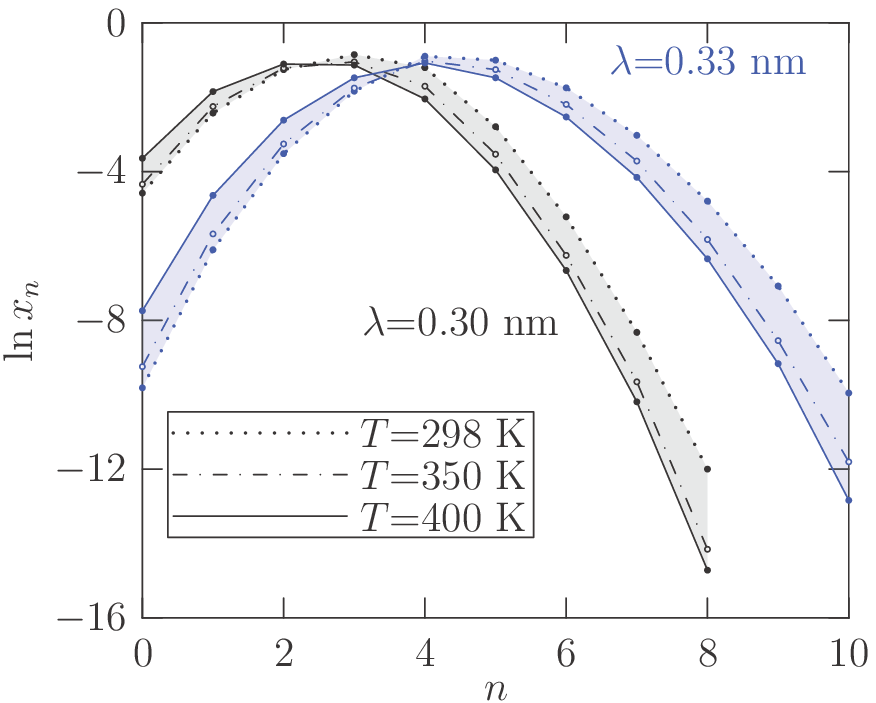}
\caption{Coordination number probabilities, $x_n$ = $p\left(n_\lambda =
n\right)$.  Note that for $\lambda$=0.30~nm, the most probable
coordination number is 2  or 3.\cite{RAHMANA:Moldsl,STILLINGERFH:Watr}
For $\lambda$=0.33~nm, the most probable coordination number is n=4. 
But $n$=4 and  $n$=5 are nearly equally probable at $T$=298~K, and $n=6$
is then 17\% of the whole.  With the TIP3P model, as $T$ is lowered from
400~K to 298~K there is a clear tendency to enhance probability of
higher coordination number cases.\label{Fig:xn} }
\end{figure}

The fluctuation contribution exhibits a broad maximum for $\lambda <$
0.29~nm, after which this contribution decreases steadily with
increasing $\lambda$ [Table~\ref{tab:results}].  Evidently water
molecules closest to the distinguished molecule,  \emph{i.e.,} those
closer  than the principal maximum of oxygen-oxygen radial distribution
function, don't contribute importantly to the fluctuations. This is
consistent with a quasi-chemical picture in which a water molecule and
its nearest neighbors have a definite structural integrity.
Fig.~\ref{Fig:xn} shows $x_n \propto K_n\rho_\mathrm{w}{}^n$; the
most probable coordination number is either 2 or 3 when
$\lambda$=0.30~nm, but it is 4 when $\lambda$=0.33~nm.
Nevertheless, the breadth of this distribution is remarkable.  The
coordination number $n$=6 is more probable than $n$=2 for the $\lambda$
= 0.33~nm~results of Fig.~\ref{Fig:xn}.

The magnitude of the individual contributions to $\mu^\mathrm{{ex}}$ are
of the same order as the net free energy; the mean binding energies are
larger than that, as are the  variance contributions in some cases.  The
variance contributions are about half as large as the mean binding
energies, with opposite sign. It is remarkable and significant,
therefore, that the net free energies at all temperatures are within
roughly 12\% of the numerically exact value computed by the
histogram-overlap method. The predicted excess entropy,
$S^{\mathrm{ex}}/nk$, (Fig.~\ref{fig:murdep}) is in error by  17\% at
the lowest temperature considered and more accurate than that at the
higher temperatures. The constant-pressure heat capacity predicted by
the model is roughly 30\% larger than the result obtained using the
numerically exact values for $S^{\mathrm{ex}}/nk$ in the same
finite-difference estimate.
%\cite{MARCUSY:Assw} 
A mean-field-like approximation that neglects
fluctuations produces useless results.

The consistent combination of the various terms is an important
observation.  For example, the packing contribution,
$\mu^{\mathrm{ex}}_{\mathrm{HS}}\left(\lambda\right)$, is sensitive to
the interaction potential energy model used, and primarily through
differences in the densities of different models at the higher
temperatures.\cite{PaschekD:Temdhh} We found that use of  values of
$\mu^{\mathrm{ex}}_{\mathrm{HS}}\left(\lambda\right)$ corresponding to a
 more accurate model of liquid water \cite{Ashbaugh:2006} spoils its
consistency with the other contributions evaluated here with the TIP3P
model, and that appreciably degrades the accuracy of the whole.

If $kT\ln{p(n_\lambda = 0)}$ is zero [Table~\ref{tab:results}], the
hard-sphere contribution
$\mu^\mathrm{{ex}}_\mathrm{{HS}}\left(\lambda\right)$ is ill-defined. As
a general matter, the sum
$\mu^\mathrm{{ex}}_\mathrm{{HS}}\left(\lambda\right)+kT\ln{p(n_\lambda =
0)}$ cannot be identified as a hard-sphere contribution.  Since these
terms have opposite signs, the net value can be zero or negative, and
those possibilities are realized in Table~\ref{tab:results}.  To define
the hard-sphere contribution more generally, we require that
$\mu^\mathrm{{ex}}_\mathrm{{HS}}\left(\lambda\right)$ be continuous as
$p(n_\lambda = 0)\rightarrow$ 1 with decreasing $\lambda$. All other
terms of Eq.~~\eqref{eq:working_eq_gaussian} will be independent of
$\lambda$ for values smaller than that, and it is natural to require
that of $\mu^\mathrm{{ex}}_\mathrm{{HS}}\left(\lambda\right)$ also.

From Fig.~\ref{fig:murdep}, we see that $\lambda >$ 0.30~nm
identifies a larger-size regime where the variation of the free energy
with $\lambda$ is not statistically significant. Although we anticipate
a decay toward the numerically exact value for $\lambda \rightarrow
\infty$, the statistical errors become unmanageable for values of
$\lambda$ much larger than 0.30~nm.   When $\lambda$ = 0.30~nm a
significant skew  in $P(\varepsilon \vert n_\lambda = 0)$ is not
observed, as already noted with
Fig.~\ref{Fig:comparison_rcut_full_T300}.

\begin{figure}[t]
\includegraphics[width=3.2in]{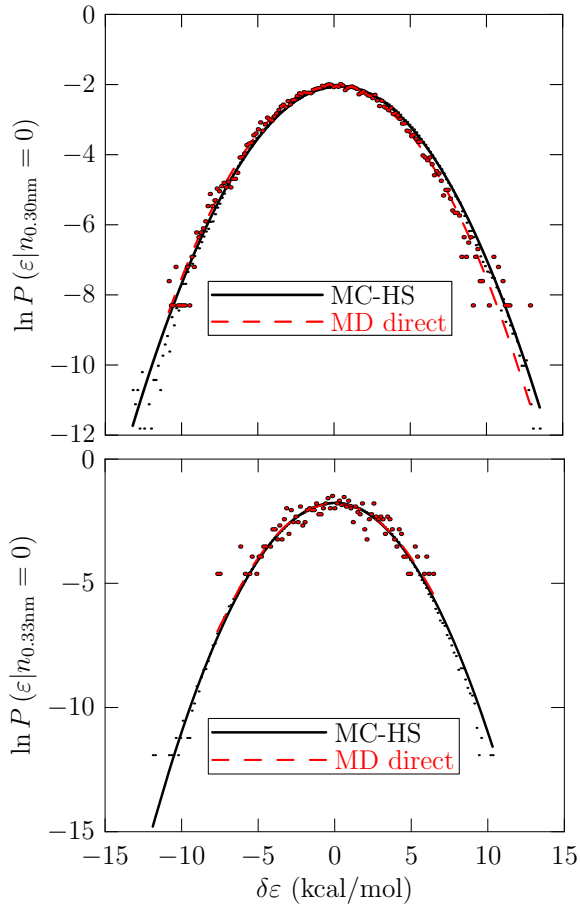}
\caption{The fine black points are estimates of $P\left(\varepsilon\vert
n_\lambda = 0\right)$ obtained from Monte Carlo simulation of TIP3P
water with one water molecule centering a hard-sphere which perfectly
repels other water oxygens from a sphere of radius $\lambda$.  (upper
panel) $\lambda$ = 0.30~nm; (lower panel) $\lambda$ = 0.33~nm.  The
larger composite dots were obtained directly from molecular dynamics
simulations with no constraint.  The solid curve is the gaussian model
for the expanded Monte Carlo data set, and confirms both the overall
accuracy of the gaussian model and the estimated conditional mean and
variance of Table I.  Note that the data are slightly less than the
gaussian model in the high-$\varepsilon$ tail.  $T$=298K.\label{Fig:pe} }
\end{figure}

\begin{figure}[t]
\includegraphics[width=3.2in]{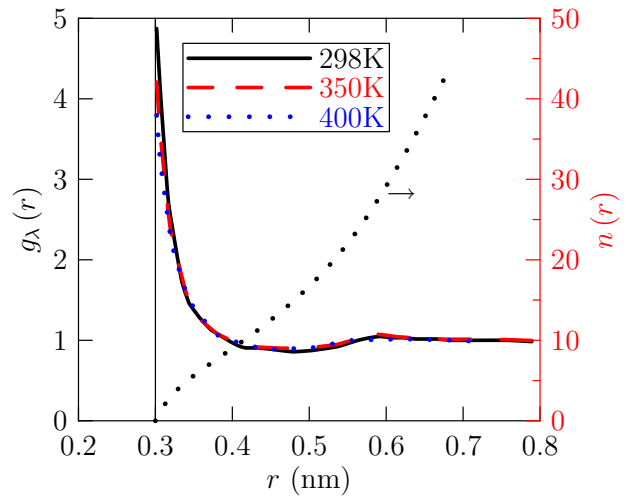}
\caption{Distribution of water oxygen atoms radially from a
distinguished water oxygen atom; evaluated with the sample corresponding
to the condition $n_\lambda = 0$, with $\lambda=$~0.30~nm, for O-atom
neighbors of the distinguished O-atom. The black, dotted curve is the
running coordination number for the $T$=298K case.\label{Fig:grlambda} }
\end{figure}

The predicted free energy $\mu^\mathrm{{ex}}$ is then distinctly above
the numerically exact value, suggesting that the gaussian model predicts
too much weight in the high-$\varepsilon$ tail. We further examined  the
high-$\varepsilon$ tail of $P\left(\varepsilon\vert n_\lambda = 0\right)$ for
larger values of $\lambda$ by carrying-out Monte Carlo calculations
under the same conditions but with one water molecule
carrying an O-centered sphere that prohibits other
O-atoms within a  radius $\lambda$.   Those results are shown in
Fig.~\ref{Fig:pe}.  The accuracy of a gaussian model
$P\left(\varepsilon\vert n_\lambda = 0\right)$ is clearly born-out, 
but the gaussian model predicts 
slightly too much probability in the extreme
high-$\varepsilon$ region particularly for $\lambda$ = 0.33~nm. The
radial distribution of outer-shell neighbors in the conditioned sample
(Fig.~\ref{Fig:grlambda}) shows that in that case the mean
water-molecule-density is structured as a thin radial shell with roughly
$n \approx$ 15 near-neighbors.

The design strategy here is that the theory should be correct in the
$\lambda\rightarrow\infty$ limit.   The accuracy of this approximate
theory should then be judged for the largest values of $\lambda$ that
are practical.  Consulting the direct investigation of the
high-$\varepsilon$ tail (Fig.~\ref{Fig:pe}), we see that the probability density
drops off by roughly $\me^{-10}$ from the mode to the
highest-$\varepsilon$ data there.   But the integrand factor of
$\me^{\beta\varepsilon}$ of Eq.~\ref{eq:derivation} grows by roughly $\me^{+20}$ across
that range; these distributions are remarkably wide even after the
conditioning.   Thus a statistical model that extends those data is
still necessary.

In the context of implementations of quasi-chemical theory, it has been
argued that the cut-off parameter  $\lambda$ may be determined on a
variational basis.  Inner- and outer-shell contributions have
countervailing effects, so where the net-results are insensitive to
local changes of $\lambda$ distinct approximations may be considered
well balanced.\cite{Beck:2006}  Here, the contributions corresponding
to the inner-shell contribution in quasi-chemical theory are obtained
numerically exactly so that argument isn't compelling.

A previous quasi-chemical analysis of liquid water,\cite{Asthagiri:2003} which utilized \emph{ab initio} molecular
dynamics (AIMD) data, provides an interesting comparison with the
present results.  In initiating analysis of AIMD, that previous effort
addressed a much more complicated case.  Several contributions had to be
estimated on the basis of slightly extraneous input, \emph{e.g.,}
dielectric model calculations and rough estimates of the effects of
outer-shell dispersion contributions.  Nevertheless, there was some
commonality in that both mean-field and fluctuation (dielectric) effects
were involved in the outer-shell contributions.  The net accuracy in
that previous case was similar to that observed for the simpler case,
more thoroughly analyzed here.  The evaluated free energy was too high
but by less than 1~kcal/mol at moderate temperatures.

Review of the theory Eq.~\ref{eq:working_eq_gaussian} identifies a
further remarkable observation.   If we follow that logic in reverse,
exploiting the fact that $\mu^{\mathrm{ex}}$ is an experimentally known
quantity, then Eq.~\ref{eq:working_eq_gaussian} provides a physically
transparent theory for $\mu^{\mathrm{ex}}_{\mathrm{HS}}\left(\lambda
\right)$ the object of central interest in theories of hydrophobic
effects.\cite{PrattLR:Molthe,Ashbaugh:2006} As a theory of hydrophobic
effects, Eq.~\ref{eq:working_eq_gaussian} offers a specific accounting
for the effects of the distinctive coordination characteristics of a
water molecule in liquid water, and a specific accounting of  the role
of longer-than-near-neighbor-ranged interactions.

\section*{Conclusions} 
Viewed from the perspective of Eq.~\ref{eq:theorem}, which is precisely
correct for liquid water models with interactions shorter-ranged that
$\lambda$, the quantitative contribution from longer-ranged interactions
for the TIP3P model is essentially identical to the experimental value
of $\mu^\mathrm{{ex}}$ for reasonable values of $\lambda\approx$
0.33~nm, independently of temperature.  If these longer-ranged
contributions were modeled as a mean-field contribution
$\left<\varepsilon\vert n_\lambda = 0\right> $ only, the error would be
nearly 100\%. On the basis of the present work, it is evident that
%Ref.~\cite{petrone:jmb04}, 
Ref.~35, 
in adopting the value $\lambda$ = 0.28~nm for
$\mu^\mathrm{{ex}}_\mathrm{{HS}} \left(\lambda\right)$, assumed that
quantitative chemical contributions reflecting short-ranged local order
were negligible, independently of the local environment.  This work then
clarifies that for bulk water $\lambda$ = 0.26~nm, rather than 0.28~nm, 
if those chemical contributions are to be negligible.  Moreover,
suitable values of $\lambda$ require additional scrutiny for different
local environments.

The present theory helps to gauge the significance of the various features of
the interactions for the molecular of liquid water.  As an example
consider the case of an interaction model with near-neighbor-ranged
interactions only. Then Eq.~\ref{eq:theorem} applies.  Suppose that the
parameters of the model are adjusted to match precisely the observed
coordination number probabilities (Fig.~\ref{Fig:xn}) for realistic
models. If the packing contribution is obtained from 
physical data\cite{HummerG:Anitm,PaschekD:Temdhh} or from simulation of realistic models
of liquid water,\cite{Ashbaugh:2006}
then it will have positive magnitudes consistent with the values of Table~\ref{tab:results}. Then we must
anticipate that the near-neighbor-ranged interaction model will yield
errors in the free  energy of roughly 100\%. Contrapositively, if the
parameters of such a model are adjusted to yield an accurate free
energies, $\mu^\mathrm{{ex}}$ 
and $\mu^\mathrm{{ex}}_\mathrm{{HS}} \left(\lambda\right)$, the results above together with the theorem Eq.~\ref{eq:theorem}
suggest that the predicted coordination number distribution ---
specifically $p\left(n_\lambda=0\right)$ = $x_0$ --- should differ
significantly from results for models with realistic longer-ranged
interactions. The implication of the analysis presented here for network
models of liquid water is that a detailed treatment of local order alone
-- {\it e.g.}, invoking specific short-ranged hydrogen-bonding
interactions that capture geometric features of the tetrahedral
structure of ice but not involving longer-ranged interactions
-- does not generally describe the range of
thermodynamic properties that distinguishes water from organic solvents.

A more specific conclusion is that the observed $P\left(\varepsilon\vert
n_\lambda = 0 \right)$ are unimodal. The gaussian model investigated
here is surprisingly realistic; the  molar excess entropies are
predicted with an accuracy of magnitude roughly $k$ (Boltzmann's
constant). The predicted free energies are more  accurate than
1~kcal/mol.  The distinctions from gaussian form are subtle quantitative
problems that have significant consequences for the thermodynamic
properties that distinguish water from many other liquids.   Those
quantitative subtleties remain to be further analyzed.

\section*{Acknowledgements}
We thank H. S. Ashbaugh for pointing-out that consistency of the
evaluated $\mu^{\mathrm{ex}}_{\mathrm{HS}}\left(\lambda \right)$ with
the other contributions could be important. This work was carried out
under the auspices of the National Nuclear Security Administration of
the U.S. Department of Energy at Los Alamos National Laboratory under
Contract No. DE-AC52-06NA25396. Financial support from the National
Science Foundation (BES0555281) is gratefully acknowledged. 
LA-UR-07-3494

%\bibliography{references}

\begin{thebibliography}{36}
\expandafter\ifx\csname natexlab\endcsname\relax\def\natexlab#1{#1}\fi
\expandafter\ifx\csname bibnamefont\endcsname\relax
  \def\bibnamefont#1{#1}\fi
\expandafter\ifx\csname bibfnamefont\endcsname\relax
  \def\bibfnamefont#1{#1}\fi
\expandafter\ifx\csname citenamefont\endcsname\relax
  \def\citenamefont#1{#1}\fi
\expandafter\ifx\csname url\endcsname\relax
  \def\url#1{\texttt{#1}}\fi
\expandafter\ifx\csname urlprefix\endcsname\relax\def\urlprefix{URL }\fi
\providecommand{\bibinfo}[2]{#2}
\providecommand{\eprint}[2][]{\url{#2}}

\bibitem[{\citenamefont{Peery and Evans}(2003)}]{PeeryTB:Assfwf}
\bibinfo{author}{\bibfnamefont{T.~B.} \bibnamefont{Peery}} \bibnamefont{and}
  \bibinfo{author}{\bibfnamefont{G.~T.} \bibnamefont{Evans}},
  \bibinfo{journal}{J. Chem. Phys.} \textbf{\bibinfo{volume}{118}},
  \bibinfo{pages}{2286 } (\bibinfo{year}{2003}).

\bibitem[{\citenamefont{Andersen}(1973)}]{ANDERSENHC:CLUEFH1}
\bibinfo{author}{\bibfnamefont{H.~C.} \bibnamefont{Andersen}},
  \bibinfo{journal}{J. Chem. Phys.} \textbf{\bibinfo{volume}{59}},
  \bibinfo{pages}{4714 } (\bibinfo{year}{1973}).

\bibitem[{\citenamefont{Andersen}(1974)}]{ANDERSENHC:CLUEFH2}
\bibinfo{author}{\bibfnamefont{H.~C.} \bibnamefont{Andersen}},
  \bibinfo{journal}{J. Chem. Phys.} \textbf{\bibinfo{volume}{61}},
  \bibinfo{pages}{4985 } (\bibinfo{year}{1974}).

\bibitem[{\citenamefont{Dahl and Andersen}(1983)}]{DAHLLW:Atat}
\bibinfo{author}{\bibfnamefont{L.~W.} \bibnamefont{Dahl}} \bibnamefont{and}
  \bibinfo{author}{\bibfnamefont{H.~C.} \bibnamefont{Andersen}},
  \bibinfo{journal}{J. Chem. Phys.} \textbf{\bibinfo{volume}{78}},
  \bibinfo{pages}{1980 } (\bibinfo{year}{1983}).

\bibitem[{\citenamefont{Aranovich et~al.}(1999)\citenamefont{Aranovich,
  Donohue, and Donohue}}]{AranovichG:Almf}
\bibinfo{author}{\bibfnamefont{G.}~\bibnamefont{Aranovich}},
  \bibinfo{author}{\bibfnamefont{P.}~\bibnamefont{Donohue}}, \bibnamefont{and}
  \bibinfo{author}{\bibfnamefont{M.}~\bibnamefont{Donohue}},
  \bibinfo{journal}{J. Chem. Phys.} \textbf{\bibinfo{volume}{111}},
  \bibinfo{pages}{2050 } (\bibinfo{year}{1999}).

\bibitem[{\citenamefont{Economou and Donohue}(1991)}]{ECONOMOUIG:Cheqap}
\bibinfo{author}{\bibfnamefont{I.~G.} \bibnamefont{Economou}} \bibnamefont{and}
  \bibinfo{author}{\bibfnamefont{M.~D.} \bibnamefont{Donohue}},
  \bibinfo{journal}{AIChE J.} \textbf{\bibinfo{volume}{37}},
  \bibinfo{pages}{1875 } (\bibinfo{year}{1991}).

\bibitem[{\citenamefont{Fleming and Gibbs}(1974{\natexlab{a}})}]{FLEMINGPD:I}
\bibinfo{author}{\bibfnamefont{P.~D.} \bibnamefont{Fleming}} \bibnamefont{and}
  \bibinfo{author}{\bibfnamefont{J.~H.} \bibnamefont{Gibbs}},
  \bibinfo{journal}{J. Stat. Phys.} \textbf{\bibinfo{volume}{10}},
  \bibinfo{pages}{157 } (\bibinfo{year}{1974}{\natexlab{a}}).

\bibitem[{\citenamefont{Fleming and Gibbs}(1974{\natexlab{b}})}]{FLEMINGPD:II}
\bibinfo{author}{\bibfnamefont{P.~D.} \bibnamefont{Fleming}} \bibnamefont{and}
  \bibinfo{author}{\bibfnamefont{J.~H.} \bibnamefont{Gibbs}},
  \bibinfo{journal}{J. Stat. Phys.} \textbf{\bibinfo{volume}{10}},
  \bibinfo{pages}{351 } (\bibinfo{year}{1974}{\natexlab{b}}).

\bibitem[{\citenamefont{Stillinger}(1975)}]{StillingerFH:Theamm}
\bibinfo{author}{\bibfnamefont{F.~H.} \bibnamefont{Stillinger}}, in
  \emph{\bibinfo{booktitle}{Adv. Chem. Phys. vol.31, ``Non-simple liquids''}},
  edited by \bibinfo{editor}{\bibfnamefont{S.}~\bibnamefont{Prigogine},
  \bibfnamefont{I.;~Rice}} (\bibinfo{publisher}{Chichester, Sussex, UK :
  Wiley}, \bibinfo{year}{1975}), pp. \bibinfo{pages}{1 -- 101}.

\bibitem[{\citenamefont{Beck et~al.}(2006)\citenamefont{Beck, Paulaitis, and
  Pratt}}]{Beck:2006}
\bibinfo{author}{\bibfnamefont{T.~L.} \bibnamefont{Beck}},
  \bibinfo{author}{\bibfnamefont{M.~E.} \bibnamefont{Paulaitis}},
  \bibnamefont{and} \bibinfo{author}{\bibfnamefont{L.~R.} \bibnamefont{Pratt}},
  \emph{\bibinfo{title}{The potential distribution theorem and models of
  molecular solutions}} (\bibinfo{publisher}{Cambridge University Press},
  \bibinfo{address}{Cambridge}, \bibinfo{year}{2006}).

\bibitem[{\citenamefont{Asthagiri et~al.}(in press
  2007)\citenamefont{Asthagiri, Ashbaugh, Piryatinski, Paulaitis, and
  Pratt}}]{CF4}
\bibinfo{author}{\bibfnamefont{D.}~\bibnamefont{Asthagiri}},
  \bibinfo{author}{\bibfnamefont{H.~S.} \bibnamefont{Ashbaugh}},
  \bibinfo{author}{\bibfnamefont{A.}~\bibnamefont{Piryatinski}},
  \bibinfo{author}{\bibfnamefont{M.~E.} \bibnamefont{Paulaitis}},
  \bibnamefont{and} \bibinfo{author}{\bibfnamefont{L.~R.} \bibnamefont{Pratt}},
  \bibinfo{journal}{J. Am. Chem. Soc.} \textbf{\bibinfo{volume}{129}},
  \bibinfo{pages}{xyz} (\bibinfo{year}{in press 2007}).

\bibitem[{\citenamefont{Ashbaugh and Pratt}(2006)}]{Ashbaugh:2006}
\bibinfo{author}{\bibfnamefont{H.~S.} \bibnamefont{Ashbaugh}} \bibnamefont{and}
  \bibinfo{author}{\bibfnamefont{L.~R.} \bibnamefont{Pratt}},
  \bibinfo{journal}{Rev.\ Mod. \ Phys.} \textbf{\bibinfo{volume}{78}},
  \bibinfo{pages}{159} (\bibinfo{year}{2006}).

\bibitem[{\citenamefont{Pratt and Asthagiri}(2007)}]{pratt07}
\bibinfo{author}{\bibfnamefont{L.~R.} \bibnamefont{Pratt}} \bibnamefont{and}
  \bibinfo{author}{\bibfnamefont{D.}~\bibnamefont{Asthagiri}}, in
  \emph{\bibinfo{booktitle}{Free Energy Calculations. Theory and Applications
  in Chemistry and Biology}} (\bibinfo{publisher}{Springer-Verlag},
  \bibinfo{address}{Berlin}, \bibinfo{year}{2007}), chap. \bibinfo{chapter}{9.
  Potential distribution methods and free energy models of molecular
  solutions}, pp. \bibinfo{pages}{323--351}.

\bibitem[{\citenamefont{Paulaitis and Pratt}(2002)}]{Paulaitis:2002}
\bibinfo{author}{\bibfnamefont{M.~E.} \bibnamefont{Paulaitis}}
  \bibnamefont{and} \bibinfo{author}{\bibfnamefont{L.~R.} \bibnamefont{Pratt}},
  \bibinfo{journal}{Adv. \ Prot.\ Chem.} \textbf{\bibinfo{volume}{62}},
  \bibinfo{pages}{283} (\bibinfo{year}{2002}).

\bibitem[{\citenamefont{Martin et~al.}(1998)\citenamefont{Martin, Hay, and
  Pratt}}]{MartinRL:Hydfiw}
\bibinfo{author}{\bibfnamefont{R.~L.} \bibnamefont{Martin}},
  \bibinfo{author}{\bibfnamefont{P.~J.} \bibnamefont{Hay}}, \bibnamefont{and}
  \bibinfo{author}{\bibfnamefont{L.~R.} \bibnamefont{Pratt}},
  \bibinfo{journal}{J. Phys. Chem. A} \textbf{\bibinfo{volume}{102}},
  \bibinfo{pages}{3565 } (\bibinfo{year}{1998}).

\bibitem[{\citenamefont{Pratt and Rempe}(1999)}]{Rempe99}
\bibinfo{author}{\bibfnamefont{L.~R.} \bibnamefont{Pratt}} \bibnamefont{and}
  \bibinfo{author}{\bibfnamefont{S.~B.} \bibnamefont{Rempe}}, in
  \emph{\bibinfo{booktitle}{Simulation and Theory of Electrostatic Interactions
  in Solution}}, edited by \bibinfo{editor}{\bibfnamefont{L.~R.}
  \bibnamefont{Pratt}} \bibnamefont{and}
  \bibinfo{editor}{\bibfnamefont{G.}~\bibnamefont{Hummer}}
  (\bibinfo{publisher}{AIP Conference Proceedings}, \bibinfo{year}{1999}), vol.
  \bibinfo{volume}{492}, pp. \bibinfo{pages}{172--201}.

\bibitem[{\citenamefont{Rempe and Pratt}(2001)}]{RempeSB:ThehnN}
\bibinfo{author}{\bibfnamefont{S.~B.} \bibnamefont{Rempe}} \bibnamefont{and}
  \bibinfo{author}{\bibfnamefont{L.~R.} \bibnamefont{Pratt}},
  \bibinfo{journal}{Fluid Phase Equilibria} \textbf{\bibinfo{volume}{183-184}},
  \bibinfo{pages}{121 } (\bibinfo{year}{2001}).

\bibitem[{\citenamefont{Rempe et~al.}(2000)\citenamefont{Rempe, Pratt, Hummer,
  Kress, Martin, and Redondo}}]{RempeSB:ThehnL}
\bibinfo{author}{\bibfnamefont{S.~B.} \bibnamefont{Rempe}},
  \bibinfo{author}{\bibfnamefont{L.~R.} \bibnamefont{Pratt}},
  \bibinfo{author}{\bibfnamefont{G.}~\bibnamefont{Hummer}},
  \bibinfo{author}{\bibfnamefont{J.~D.} \bibnamefont{Kress}},
  \bibinfo{author}{\bibfnamefont{R.~L.} \bibnamefont{Martin}},
  \bibnamefont{and} \bibinfo{author}{\bibfnamefont{A.}~\bibnamefont{Redondo}},
  \bibinfo{journal}{J. Am. Chem. Soc.} \textbf{\bibinfo{volume}{122}},
  \bibinfo{pages}{966 } (\bibinfo{year}{2000}).

\bibitem[{\citenamefont{Asthagiri et~al.}(2004)\citenamefont{Asthagiri, Pratt,
  Paulaitis, and Rempe}}]{AsthagiriD:Hydsaf}
\bibinfo{author}{\bibfnamefont{D.}~\bibnamefont{Asthagiri}},
  \bibinfo{author}{\bibfnamefont{L.~R.} \bibnamefont{Pratt}},
  \bibinfo{author}{\bibfnamefont{M.~E.} \bibnamefont{Paulaitis}},
  \bibnamefont{and} \bibinfo{author}{\bibfnamefont{S.~B.} \bibnamefont{Rempe}},
  \bibinfo{journal}{J. Am. Chem. Soc.} \textbf{\bibinfo{volume}{126}},
  \bibinfo{pages}{1285 } (\bibinfo{year}{2004}).

\bibitem[{\citenamefont{Asthagiri
  et~al.}(2003{\natexlab{a}})\citenamefont{Asthagiri, Pratt, and
  Ashbaugh}}]{AsthagiriD:Abshfe}
\bibinfo{author}{\bibfnamefont{D.}~\bibnamefont{Asthagiri}},
  \bibinfo{author}{\bibfnamefont{L.~R.} \bibnamefont{Pratt}}, \bibnamefont{and}
  \bibinfo{author}{\bibfnamefont{H.~S.} \bibnamefont{Ashbaugh}},
  \bibinfo{journal}{J. Chem. Phys.} \textbf{\bibinfo{volume}{119}},
  \bibinfo{pages}{2702 } (\bibinfo{year}{2003}{\natexlab{a}}).

\bibitem[{\citenamefont{Asthagiri and Pratt}(2003)}]{AsthagiriD:QuasB2}
\bibinfo{author}{\bibfnamefont{D.}~\bibnamefont{Asthagiri}} \bibnamefont{and}
  \bibinfo{author}{\bibfnamefont{L.~R.} \bibnamefont{Pratt}},
  \bibinfo{journal}{Chem. Phys. Letts.} \textbf{\bibinfo{volume}{371}},
  \bibinfo{pages}{613 } (\bibinfo{year}{2003}).

\bibitem[{\citenamefont{Asthagiri
  et~al.}(2003{\natexlab{b}})\citenamefont{Asthagiri, Pratt, Kress, and
  Gomez}}]{AsthagiriD:ThehsH}
\bibinfo{author}{\bibfnamefont{D.}~\bibnamefont{Asthagiri}},
  \bibinfo{author}{\bibfnamefont{L.~R.} \bibnamefont{Pratt}},
  \bibinfo{author}{\bibfnamefont{J.~D.} \bibnamefont{Kress}}, \bibnamefont{and}
  \bibinfo{author}{\bibfnamefont{M.~A.} \bibnamefont{Gomez}},
  \bibinfo{journal}{Chem. Phys. Letts.} \textbf{\bibinfo{volume}{380}},
  \bibinfo{pages}{530 } (\bibinfo{year}{2003}{\natexlab{b}}).

\bibitem[{\citenamefont{Rempe et~al.}(2004)\citenamefont{Rempe, Asthagiri, and
  Pratt}}]{RempeSB:Innsda}
\bibinfo{author}{\bibfnamefont{S.~B.} \bibnamefont{Rempe}},
  \bibinfo{author}{\bibfnamefont{D.}~\bibnamefont{Asthagiri}},
  \bibnamefont{and} \bibinfo{author}{\bibfnamefont{L.~R.} \bibnamefont{Pratt}},
  \bibinfo{journal}{PCCP} \textbf{\bibinfo{volume}{6}}, \bibinfo{pages}{1966 }
  (\bibinfo{year}{2004}).

\bibitem[{\citenamefont{Pratt and Laviolette}(1998)}]{PrattLR:Quatal}
\bibinfo{author}{\bibfnamefont{L.~R.} \bibnamefont{Pratt}} \bibnamefont{and}
  \bibinfo{author}{\bibfnamefont{R.~A.} \bibnamefont{Laviolette}},
  \bibinfo{journal}{Mol. Phys.} \textbf{\bibinfo{volume}{94}},
  \bibinfo{pages}{909 } (\bibinfo{year}{1998}).

\bibitem[{\citenamefont{Rahman and Stillinger}(1971)}]{RAHMANA:Moldsl}
\bibinfo{author}{\bibfnamefont{A.}~\bibnamefont{Rahman}} \bibnamefont{and}
  \bibinfo{author}{\bibfnamefont{F.~H.} \bibnamefont{Stillinger}},
  \bibinfo{journal}{J. Chem. Phys.} \textbf{\bibinfo{volume}{55}},
  \bibinfo{pages}{3336 } (\bibinfo{year}{1971}).

\bibitem[{\citenamefont{Stillinger}(1980)}]{STILLINGERFH:Watr}
\bibinfo{author}{\bibfnamefont{F.~H.} \bibnamefont{Stillinger}},
  \bibinfo{journal}{Science} \textbf{\bibinfo{volume}{209}},
  \bibinfo{pages}{451 } (\bibinfo{year}{1980}).

\bibitem[{\citenamefont{Paliwal et~al.}(2006)\citenamefont{Paliwal, Asthagiri,
  Pratt, Ashbaugh, and Paulaitis}}]{PaliwalA.:Anamp}
\bibinfo{author}{\bibfnamefont{A.}~\bibnamefont{Paliwal}},
  \bibinfo{author}{\bibfnamefont{D.}~\bibnamefont{Asthagiri}},
  \bibinfo{author}{\bibfnamefont{L.~R.} \bibnamefont{Pratt}},
  \bibinfo{author}{\bibfnamefont{H.~S.} \bibnamefont{Ashbaugh}},
  \bibnamefont{and} \bibinfo{author}{\bibfnamefont{M.~E.}
  \bibnamefont{Paulaitis}}, \bibinfo{journal}{J. Chem. Phys.}
  \textbf{\bibinfo{volume}{124}} (\bibinfo{year}{2006}).

\bibitem[{\citenamefont{Paliwal}(2005)}]{Paliwal:thesis}
\bibinfo{author}{\bibfnamefont{A.}~\bibnamefont{Paliwal}}, Ph.D. thesis,
  \bibinfo{school}{Johns Hopkins University} (\bibinfo{year}{2005}).

\bibitem[{\citenamefont{Andaloro and
  Sperandeo-Mineo}(1990)}]{AndaloroG.:MonCsh}
\bibinfo{author}{\bibfnamefont{G.}~\bibnamefont{Andaloro}} \bibnamefont{and}
  \bibinfo{author}{\bibfnamefont{R.~M.} \bibnamefont{Sperandeo-Mineo}},
  \bibinfo{journal}{Eur. J. Phys.} \textbf{\bibinfo{volume}{11}},
  \bibinfo{pages}{275 } (\bibinfo{year}{1990}).

\bibitem[{\citenamefont{Haymet et~al.}(1996)\citenamefont{Haymet, Silverstein,
  and Dill}}]{HaymetADJ:Hydr'e}
\bibinfo{author}{\bibfnamefont{A.~D.~J.} \bibnamefont{Haymet}},
  \bibinfo{author}{\bibfnamefont{K.~A.~T.} \bibnamefont{Silverstein}},
  \bibnamefont{and} \bibinfo{author}{\bibfnamefont{K.~A.} \bibnamefont{Dill}},
  \bibinfo{journal}{Faraday Discussions} \textbf{\bibinfo{volume}{103}},
  \bibinfo{pages}{117 } (\bibinfo{year}{1996}).

\bibitem[{\citenamefont{Silverstein et~al.}(1998)\citenamefont{Silverstein,
  Haymet, and Dill}}]{SilversteinKAT:Asmw}
\bibinfo{author}{\bibfnamefont{K.~A.~T.} \bibnamefont{Silverstein}},
  \bibinfo{author}{\bibfnamefont{A.~D.~J.} \bibnamefont{Haymet}},
  \bibnamefont{and} \bibinfo{author}{\bibfnamefont{K.~A.} \bibnamefont{Dill}},
  \bibinfo{journal}{J. Am. Chem. Soc.} \textbf{\bibinfo{volume}{120}},
  \bibinfo{pages}{3166 } (\bibinfo{year}{1998}).

\bibitem[{\citenamefont{Paschek}(2004)}]{PaschekD:Temdhh}
\bibinfo{author}{\bibfnamefont{D.}~\bibnamefont{Paschek}}, \bibinfo{journal}{J.
  Chem. Phys.} \textbf{\bibinfo{volume}{120}}, \bibinfo{pages}{6674 }
  (\bibinfo{year}{2004}).

\bibitem[{\citenamefont{Asthagiri
  et~al.}(2003{\natexlab{c}})\citenamefont{Asthagiri, Pratt, and
  Kress}}]{Asthagiri:2003}
\bibinfo{author}{\bibfnamefont{D.}~\bibnamefont{Asthagiri}},
  \bibinfo{author}{\bibfnamefont{L.~R.} \bibnamefont{Pratt}}, \bibnamefont{and}
  \bibinfo{author}{\bibfnamefont{J.~D.} \bibnamefont{Kress}},
  \bibinfo{journal}{Phys. Rev. E} \textbf{\bibinfo{volume}{68}},
  \bibinfo{pages}{041505 } (\bibinfo{year}{2003}{\natexlab{c}}).

\bibitem[{\citenamefont{Pratt}(2002)}]{PrattLR:Molthe}
\bibinfo{author}{\bibfnamefont{L.~R.} \bibnamefont{Pratt}},
  \bibinfo{journal}{Ann. Rev. Phys. Chem.} \textbf{\bibinfo{volume}{53}},
  \bibinfo{pages}{409 } (\bibinfo{year}{2002}).

\bibitem[{\citenamefont{Hummer et~al.}(1996)\citenamefont{Hummer, Garde,
  Garc\'{i}a, Pohorille, and Pratt}}]{HummerG:Anitm}
\bibinfo{author}{\bibfnamefont{G.}~\bibnamefont{Hummer}},
  \bibinfo{author}{\bibfnamefont{S.}~\bibnamefont{Garde}},
  \bibinfo{author}{\bibfnamefont{A.~E.} \bibnamefont{Garc\'{i}a}},
  \bibinfo{author}{\bibfnamefont{A.}~\bibnamefont{Pohorille}},
  \bibnamefont{and} \bibinfo{author}{\bibfnamefont{L.~R.} \bibnamefont{Pratt}},
  \bibinfo{journal}{Proc. Natl. Acad. Sci. USA} \textbf{\bibinfo{volume}{93}},
  \bibinfo{pages}{8951 } (\bibinfo{year}{1996}).

\bibitem[{\citenamefont{Rowlinson and Swinton}(1982)}]{Rowlinson}
\bibinfo{author}{\bibfnamefont{J.~S.} \bibnamefont{Rowlinson}}
  \bibnamefont{and} \bibinfo{author}{\bibfnamefont{F.~L.}
  \bibnamefont{Swinton}}, \emph{\bibinfo{title}{Liquids and Liquid Mixtures}}
  (\bibinfo{publisher}{Butterworths}, \bibinfo{address}{NY},
  \bibinfo{year}{1982}).

\end{thebibliography}

\clearpage

\clearpage

\end{document}